\documentclass[preprint,proceedings]{rmaa}

\usepackage{paralist}

\SetYear{2005}
\SetConfTitle{IX$^{th}$ Tex-Mex Conference}

\title{Oxygen Recombination Line Abundances in Gaseous Nebulae}

\author{
  Antonio Peimbert\altaffilmark{1} and
  Manuel Peimbert\altaffilmark{1}}

\altaffiltext{1}{Instituto de Astronom\'\i{}a, U.N.A.M.
  M\'exico, D.F., M\'exico.}

\shortauthor{Peimbert \& Peimbert}

\shorttitle{Recombination Line Abundances}

\fulladdresses{
\item A. Peimbert and M. Peimbert: Instituto de Astronom\'\i{}a, 
  Universidad Nacional Aut\'o\-no\-ma de M\'exico, 
  Apartado Postal 70-264, 04510
  M\'exico, D.F., M\'exico (\email{antonio@astroscu.unam.mx},
  \email{peimbert@astroscu.unam.mx})}

\listofauthors{A. Peimbert, \& M. Peimbert}

\indexauthor{Peimbert, A.}
\indexauthor{Peimbert, M.}

\abstract{The determination of the heavy element abundances from giant
extragalactic \ion{H}{2} regions has been generally based on
collisionally excited lines. We will discuss the reasons to study the
characteristics of recombination lines, and then use these lines to
determine chemical abundances. Of these lines the oxygen (specifically
the \ion{O}{2}) lines are the most important; and, of them, the lines
of multiplet 1 of \ion{O}{2} are the most accessible. It has often been
assumed that by measuring the intensity of a single line within a
multiplet the intensities of all the lines in the multiplet can be
determined; in recent studies we have found that the intensity ratios
of lines within a multiplet can depend on density; we will present 
empirical density-intensity relationships for multiplet 1 based on recent
observations of \ion{H}{2} regions and planetary nebulae. From observations
deof \ion{H}{2} regions we find that the critical density for collisional
redistribution of the multiplet 1 \ion{O}{2} recombination lines amounts
to $2800 \pm 500$ cm$^{-3}$.
We point out that the O/H recombination abundances of \ion{H}{2} regions 
in the solar vicinity are in excellent agreement with the O/H solar value,
while the abundances derived from collisionally excited lines are not.
We present a calibration of Pagel's method in the 8.2 $<$ 12 + log O/H 
$<$ 8.8 range based on O recombination lines.}

\resumen{La determinaci\'on de abundancias de elementos pesados es regiones 
\ion{H}{2} extragal\'acticas tradicionalmente se ha basado en l\'\i{}neas 
colisionalemente excitadas. 
Discutimos la importancia de estudiar las l\'\i{}neas de 
recombinaci\'on para determinar las abundancias de los elementos
pesados en nebulosas gaseosas. Las l\'\i{}neas de recombinaci\'on m\'as
importantes de los elementos pesados son las de ox\'\i{}geno, y de estas
las del multiplete 1 de \ion{O}{2} son las m\'as f\'aciles de observar. Con
frecuencia se supone que a partir de la intensidad de una l\'\i{}nea de un
multiplete se puede encontrar la intensidad de todas las dem\'as; en estudios
recientes hemos encontrado que los cocientes de intensidades entre
l\'\i{}neas del mismo multiplete dependen de la densidad; presentamos
relaciones emp\'\i{}ricas entre la densidad y la intensidad para el
multiplete 1 del O basados en observaciones recientes de regiones
\ion{H}{2} y 
nebulosas planetarias. A partir  las observaciones de las regiones
\ion{H}{2} encontramos que la densidad cr\'\i{}tica asociada a la
redistribucion colisional de las intensidades de las l\'\i{}neas del
multiplete 1 de \ion{O}{2} es de $2800 \pm 500$ cm$^{-3}$. Hacemos notar
que las abudancias de O obtenidas a partir de l\'\i{}neas de recombinaci\'on
muestran un acuerdo excelente con las abundancias solares, mientras
que las abundancias de O estimadas a partir de l\'\i{}neas de O excitadas
colisionalmente est\'an en desacuerdo con las abundancias solares.
Presentamos una calibraci\'on del m\'etodo de Pagel para el intervalo
8.2 $<$ 12 + log O/H $<$ 8.8 basada en l\'\i{}neas de recombinaci\'on del O.}

\addkeyword{Galaxies: Abundances}
\addkeyword{H~II regions: Abundances}
\addkeyword{Planetary Nebulae: Abundances}

\begin{document}
\maketitle

\section{Introduction}
\label{sec:intro}

There are many observations that indicate the presence of large 
temperature variations in gaseous nebulae. In the case of chemically 
homogeneous nebulae it can be shown that recombination lines provide
us with a better indication of the abundances than collisionally
excited lines. In this short review we discuss the information
provided by the \ion{O}{2} recombination lines on the determination of
O/H values in gaseous nebulae. In Section 2 we discuss the derivation
of the O/H abundances and their dependence in the electron
temperature and the electron density. In Section 3 we discuss
two independent methods to estimate the O/H value in the ISM of the
solar vicinity. In Section 4 we discuss Pagel's method to determine
the O/H abundances in extragalactic nebulae based on the [\ion{O}{2}]
and [\ion{O}{3}] nebular excitation lines and we calibrate it, for the
first time based on \ion{O}{2} and \ion{H}{1} recombination lines. In Section 5
we present the conclusions.

\section{Oxygen Abundances Derived from Recombination Lines}
\label{sec:ORL}

Peimbert, Storey, \& Torres-Peimbert (1993), based on the recombination
coefficients for \ion{O}{2} lines computed by Storey (1994), were the first 
to determine O/H values for gaseous nebulae. The temperature dependence
of the \ion{O}{2} lines is relatively weak and very similar to that of the
\ion{H}{1} lines, therefore the O$^{++}$/H$^+$ ratios are independent of the
electron temperature. Alternatively the O$^{++}$/H$^+$ ratios derived
from collisionally excited lines do depend strongly on the temperature
(e. g.: Peimbert 1967, Peimbert \& Costero 1969, Peimbert et al. 2004).
In \ion{H}{2} regions the recombination lines typically yield abundances
higher than the optical collisionally excited lines by factors in the 
2 to 3 range. The classical definition of the mean temperature
square, $t^2$, is given by Peimbert (1967) and recent discussions on the
presence of temperature variations in gaseous nebulae have been presented 
by Torres-Peimbert 
\& Peimbert (2003), Ruiz et al. (2003), and Peimbert et al. (2004).

\begin{figure}[!t]
  \includegraphics[angle=-90,width=\columnwidth]{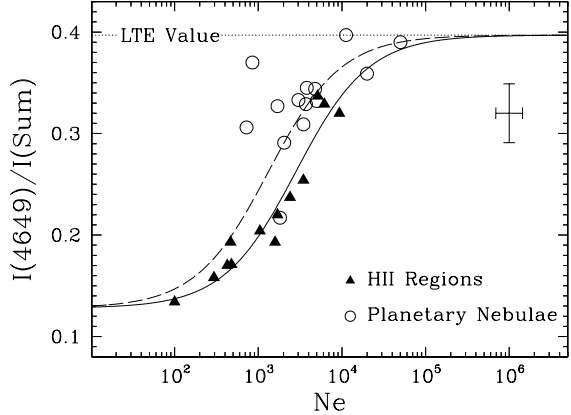}
  \caption{Intensity ratio of the \ion{O}{2} multiplet 1 line $\lambda 4649$ 
    relative to the sum of the intensities of the 8 lines of the multiplet 
    vs. the electron density derived from forbidden line ratios. 
    The solid line represents the best fit to the \ion{H}{2} region data, 
    see eq. 4; the dashed line represents a fit to most of the PNe; 
    the dotted line represents where the points would lie if there was LTE.}
  \label{fig:4649}
\end{figure}

Ruiz et al. (2003) have found that the levels where the \ion{O}{2} recombination
lines of multiplet 1 originate are not in LTE, therefore to obtain
the correct O$^{++}$/H$^+$ value it is necessary to observe the eight
lines of the multiplet. For those cases where not all of the lines
of the multiplet are observed (due to faintness of some of the lines
or due to blending produced by the low spectral resolution used) Ruiz et al. 
(2003) and Peimbert, Peimbert, \& Ruiz (2005) have presented a preliminary
set of equations of the multiplet as a function of the critical
density. Based on additional observations in what follows we will revise
those equations.
 
In Figures 1 and 2 we present additional observations of \ion{H}{2} regions
and PNe to those presented by Ruiz et al. (2003) and Peimbert et al.
(2005). We have added the \ion{H}{2} region values presented by Tsamis et al.
(2003a) for LMC11B and by Esteban et al. (2004, 2005)
and Garc\'{\i}a-Rojas et al. (2004, 2005) for Orion, M8, M17, NGC 3576, and
S 311. We have also added the PNe values presented by Tsamis et al.
(2003b) for NGC 2022, NGC 3132, NGC 3234, NGC 5882, IC 4191, and
IC 4406 and we have deleted the value for IC 4997. From the best fits to the
\ion{H}{2} regions in Figures 1 and 2 we have obtained the following
equations:
\begin{equation}
\left[ \frac{I(4651+74)}{I({\rm sum})} \right]_{obs} =
0.101 + \frac{0.128 \pm 0.010}{ \left[ 1 + N_e({\rm FL})/2800 \right] },
\label{e4651}
\end{equation}
\begin{equation}
\left[ \frac{I(4639+62+96)}{I({\rm sum})} \right]_{obs} =
0.201 + \frac{0.190 \pm 0.010}{ \left[ 1 + N_e({\rm FL})/2800 \right] },
\label{e4639}
\end{equation}
\begin{equation}
\left[ \frac{I(4642+76)}{I({\rm sum})} \right]_{obs} =
0.301 - \frac{0.049 \pm 0.010}{ \left[ 1 + N_e({\rm FL})/2800 \right] },
\label{e4642}
\end{equation}
and
\begin{equation}
\left[ \frac{I(4649)}{I({\rm sum})} \right]_{obs} =
0.397 - \frac{0.269 \pm 0.010}{ \left[ 1 + N_e({\rm FL})/2800 \right] },
\label{e4649}
\end{equation}
where the critical density, $N_c$(\ion{H}{2} regions)$ = 2800 \pm 500 $ cm$^{-3}$ 
was obtained from the observed \ion{O}{2} line intensities and the $N_e$ values 
derived from collisionally excited lines, mainly those of [\ion{Cl}{3}].

\begin{figure}[!t]
  \includegraphics[angle=-90,width=\columnwidth]{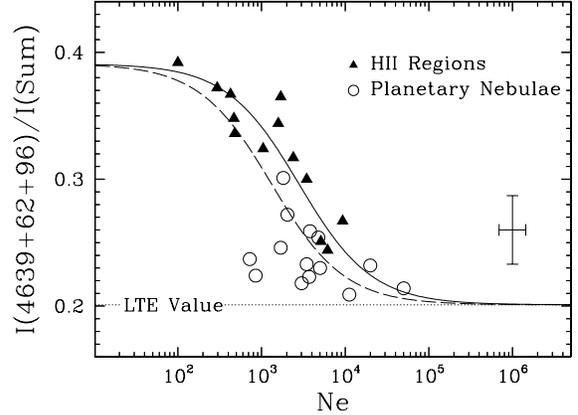}
  \caption{Same as Figure 1, but for the ratio of $\lambda \lambda 4639+62+96$ 
  relative to the sum of the lines of the multiplet.}
  \label{fig:4662}
\end{figure}

Similarly from the best fit to the PNe in Figures 1 and 2 we have obtained
a critical density, $N_c$(PNe)$ = 1325 \pm 300 $ cm$^{-3}$.

The difference between $N_c$(\ion{H}{2} regions) and $N_c$(PNe) probably indicates
that the density variations present in PNe are more significant than
those present in \ion{H}{2} regions. For the density fluctuations to be
effective in explaining this difference a significant fraction of the
emission measure should correspond to densities higher than the \ion{O}{2}
critical density and a significant fraction to densities smaller
than the critical density.

In Table 1, we present the density derived from the forbidden line
data for PNe and the density derived from the \ion{O}{2} recombination
line intensities and equations 2 and 4. The large differences
for most of the objects indicate large density fluctuations
in the 1,000 to 10,000 cm$^{-3}$ range. This fluctuations will
produce underestimations in the heavy element abundances derived
from IR collisionally excited lines if they are not taken into account.

The critical density to reach LTE for multiplet 1 of \ion{O}{2},
$N_c$(Atomic Physics) should be estimated from atomic physics
computations to see if it agrees with the one derived from
observations of \ion{H}{2} regions. If $N_c$(Atomic Physics) turns out
to be higher than $N_c$(\ion{H}{2} regions), it would indicate
the presence of significant temperature and density variations
in \ion{H}{2} regions.

\begin{table}[!t]\centering
  \setlength{\tabnotewidth}{\columnwidth}
  \tablecols{5}
  \caption{Planetary Nebula Densities} \label{tab:PNne}
  \begin{tabular}{lr@{}lr@{}l}
    \hline
    \hline
    {\small Object} & \multicolumn{2}{c}{\small Forbidden Lines} 
& \multicolumn{2}{c}{\small \ion{O}{2} multiplet 1 \tablenotemark{a}} \\
    \hline
    NGC 3132\hspace{.3cm}\null&\null\hspace{.4cm}720 & $\pm300$\hspace{.4cm}\null
               &\null\hspace{.4cm}7500 & ${+4000}\atop{-2000}$\hspace{.4cm}\null \\
    NGC 2022   &  850 & ${+1000}\atop{-500}$  & 23000 & ${+100000}\atop{-10000}$ \\
    M1-42      & 1690 & ${+600}\atop{-400}$   &  8500 &   ${+6000}\atop{-3000}$  \\
    NGC 3234   & 1800 & ${+1400}\atop{-700}$  &  1750 &   ${+1000}\atop{-700}$   \\
    NGC 5307   & 2040 & ${+1350}\atop{-1050}$ &  4500 &   ${+2250}\atop{-1250}$  \\
    NGC 7009   & 3000 & $\pm900$              & 15000 &  ${+15000}\atop{-6000}$  \\
    IC 4406    & 3500 & $\pm2000$             &  7500 &   ${+7000}\atop{-3000}$  \\
    NGC 5882   & 3700 & $\pm1500$             & 12000 &  ${+10000}\atop{-4000}$  \\
    NGC 6153   & 3830 & $\pm800$              &  8500 &   ${+7500}\atop{-3000}$  \\
    M2-36      & 4830 & $\pm1000$             &  9000 &  ${+10000}\atop{-4000}$  \\
    NGC 6543   & 5000 & $\pm1000$             & 11000 &  ${+15000}\atop{-4500}$  \\
    \hline
    \hline
    \tabnotetext{a}{Based on the \ion{H}{2} regions density calibration 
    of the \ion{O}{2} multiplet 1, average of the values derived from 
    figures 1 and 2.}
  \end{tabular}
\end{table}

As we just saw above, from the observed line intensities of a given
object and equations 1-4 it
is possible to determine the average density where the
\ion{O}{2} recombination lines originate. In the
presence of strong chemical inhomogeneities, with high density
clumps made of H-poor material embedded in low density regions of H-rich 
material, the density derived from equations 1 - 4 will be higher than
the density derived from fitting the ratio of the higher Balmer lines to
H$\beta$ or H$\alpha$.

\section{The O/H Value in the Solar Vicinity}
\label{sec:SV}

There are two independent methods to determine the O/H ratio in the
ISM of the solar vicinity: a) from the solar ratio by 
Asplund, Grevesse, \& Sauval (2005), 
that amounts to 12 + log(O/H) = 8.66, and taking into account
the increase of the O/H ratio due to galactic chemical evolution since
the Sun was formed, that according to state of the art chemical
evolution models of the Galaxy amounts to 0.13~dex (e.g. Carigi et
al. 2005), we obtain an O/H value of 8.79 dex; and b) from the \ion{H}{2}
regions O/H
value for the solar vicinity by Esteban et al. (2005) based on the O/H
galactic gradient
determined from \ion{O}{2} recombination lines, that amounts to 8.77 dex, in
excellent agreement with the value based on the solar abundance.

In the previous comparison we are assuming that the solar abundances 
are representative of the abundances of the ISM solar vicinity when
the Sun was formed. There are two other determinations of the present
O/H value in the ISM of the solar vicinity that can be made from observations of
F and G stars of the solar vicinity. According to Allende Prieto et al.
(2004) the Sun appears deficient by roughly 0.1 dex in O, Si, Ca, Sc, Ti,
Y, Ce, Nd, and Eu, compared with its immediate neighbors with similar iron
abundances, the probable reason for this 
difference is that the Sun is older than the comparison stars,
by adopting the O/H value of the comparison stars we obtain a value of
12 + log O/H = 8.76 dex. A similar result is obtained
by Bensby \& Feltzing (2005) that find that the most O-rich thin-disk F and
G dwarfs have [O/H] $\sim$ 0.15 by adopting their value for the present day
ISM of the solar vicinity we find 12 + O/H = 8.81 dex. Both results in
excellent agreement with the O/H value derived from O recombination
lines in \ion{H}{2} regions of the solar vicinity.

\section{Calibration of Pagel's Method to Derive Oxygen Abundances}
\label{sec:Pagel}

The difficulty of measuring $I(\lambda 4363)$ (or any other direct temperature
indicator) led Pagel et al. (1979) to propose an empirical method based on 
the ratio
of the nebular oxygen lines to $I({\rm H}\beta)$, $R_{23} \equiv 
I([$\ion{O}{2}$]\lambda 3727 + [$\ion{O}{3}$]\lambda\lambda 4959, 5007)/ I({\rm H}\beta)$,
to determine the O/H ratio in giant extragalactic \ion{H}{2} regions.

There are four different options to calibrate O/H versus $R_{23}$: a) from
photoionization models where the observed 
$I([$\ion{O}{2}$]\lambda 3727)/I({\rm H}\beta)$ and the
$I([$\ion{O}{3}$]\lambda\lambda 4959, 5007)/ I({\rm H}\beta)$ values are 
matched with those
predicted by the models, b) from abundances derived from the observed 
$I([$\ion{O}{2}$]\lambda 3727)/I({\rm H}\beta)$ and the
$I([$\ion{O}{3}$]\lambda\lambda 4959, 5007)/ I({\rm H}\beta)$ values
and the observed $T_e(4363/5007)$ under the assumption that $t^2$ = 0.00, 
c) from O abundances
derived from supergiant stars, and d) from O abundances derived
from recombination lines.

\subsection{Photoionization models}

This calibration is based on photoionization models where O/H is an input of
the models. Calibrations based on this option have been presented by many
authors (e. g.: McCall, Rybski, \& Shields 1985; Dopita, \& Evans 1986;
McGaugh 1991; Zaritsky, Kennicutt \& Huchra 1994; Kewley, \& Dopita 2002;
Kobulnicky, \& Kewley 2004). This 
calibration depends on
the quality of the models. A good model should include the gaseous density 
distribution for the nebula and for the ionizing
cluster: an initial mass
function, the time elapsed since the beginning of the star formation, and
a star formation rate.

The photoionization models not yet include all the physical processes needed to
reproduce all the ratios observed in real nebulae. For example 
they do not include the possible presence of stellar winds due to WR stars
nor the possible presence of supernova remnants and related shocks. From a study 
of NGC 604, a giant extragalactic \ion{H}{2} region in M33, Yang et al. (1996)
conclude that the velocity width of the H$\alpha$ line consists of equal
contributions from thermal broadening, stellar winds and SNRs, and gravity.
Even the best
photoionization models, those tailored to fit I~Zw~18, NGC~2363, and NGC~346,
predict $T_e(4363/5007)$ values smaller than observed (Stasi\'nska \& Schaerer 
1999; Luridiana, Peimbert, \& Leitherer 1999, and Rela\~no, Peimbert, 
\& Beckman 2002), probably indicating the need for additional heating sources.
The photoionization models typically predict $t^2 \approx 0.005$, values 
considerably smaller than those derived from observations that are typically 
in the 0.02 $< t^2 < 0.06$
range.

\subsection{Observations of $R_{23}$ and $T_e(4363/5007)$}

This calibration is based on adjusting the observed $R_{23}$ values with
the abundances derived from $T_e(4363/5007)$ under the assumption that $t^2 =
0.00$. These calibrations depend strongly on the temperature structure of the
nebulae and underestimate the O/H values by factors of about 2 to 3 because,
as mentioned before, $t^2$ is typically in the 0.02 to 0.06 range.

There are significant differences between the calibrations of Pagel's method
based on models (e. g. McCall et al. 1985;
Dopita \& Evans 1986; McGaugh 1991) and the calibrations based on
observations and $T_e(4363/5007)$ (e. g. Edmunds \& Pagel 1984; Torres-Peimbert,
Peimbert, \& Fierro 1989; 
Pilyugin 2000, 2003: Castellanos, D\'{\i}az, \& Terlevich 2002). The
differences in the O/H values are in the $0.2$ - $0.4$ dex range and could
be due mainly to the presence of temperature inhomogeneities over the observed
volume (e. g. Campbell 1988; Torres-Peimbert et al. 1989; McGaugh 1991;
Roy et al. 1996; Luridiana et al. 1999; Kobulnicky, Kennicutt, \& Pizagno 1999). 
These 
differences need to be sorted out if we want to obtain absolute accuracies in 
O/H of the order of 0.1 dex or better.

\subsection{O recombination lines}

\begin{figure}[!t]
  \includegraphics[width=\columnwidth]{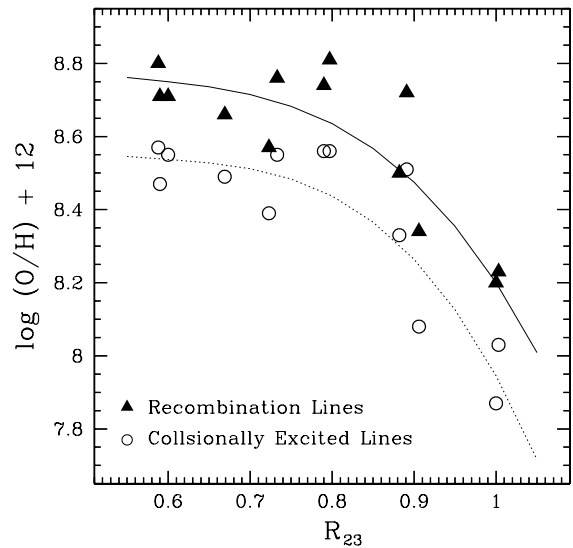}
  \caption{Pagel's R$_{23}$ method calibration --- using abundances 
    determined with recombination lines (solid line) and abundances 
    determined with collisionally excited lines (dotted line).}
  \label{fig:pagel}
\end{figure}

We have discussed previously the need to calibrate Pagel's
method with O recombination lines (Peimbert \& Peimbert 2003);
in Figure 3 we present a calibration of Pagel's method based on recombination
line observations of galactic and extragalactic \ion{H}{2} regions 
obtained by the following authors: Esteban et al. (2002, 2004, 2005),
Peimbert (2003), Tsamis et al. (2003a), Garc\'{\i}a-Rojas et al. (2004, 2005),
and Peimbert et al. (2005).
Also in this figure we present the abundances derived from the O collisionally
excited lines and $T_e(4363/5007)$ under the assumption of constant temperature,
i. e. $t^2$ = 0.00. The average difference between both methods amounts to
0.21 dex. Notice that we are presenting for both methods the gaseous abundances
without correction for the fraction of O embedded in dust grains.

We consider that the option to calibrate Pagel's method based on the O
recombination lines is superior to the one based on fitting the 5007 and
3727 O lines to those predicted by photoionization models because even the best
available models are not yet able to reproduce all the observed emission line
ratios. The O recombination method is also better than the option based on 
the observationally determined
$T_e(4363/5007)$ because the abundances derived from the nebular lines
and $T_e(4363/5007)$ are very sensitive to the $t^2$ value while the O/H values
derived from recombination lines are independent from it.

\section{Conclusions}
\label{sec:DC}

The \ion{O}{2} relative line intensities from multiplet 1 might deviate
from the LTE predictions. If that is the case equations 1 - 4 should
be used to estimate the intensities of the unobserved lines of the multiplet.

If the density derived from equations 1 - 4 is larger than that derived
from the high $n$ Balmer lines, this would indicate that the \ion{O}{2} lines originate
in high density H-poor gas embedded in a low-density H-rich gas.

The $N_c$(PNe) that we derive from the multiplet of \ion{O}{2} and forbidden 
lines is considerably smaller than the $N_c$(\ion{H}{2} regions) one, which 
implies that PNe show larger density fluctuations in the 1,000 to 10,000 cm$^{-3}$
range than \ion{H}{2} regions.

The typical densities for PNe from the \ion{O}{2} multiplet 1
calibrated based on \ion {H}{2} regions (equations 2 and 4) are about
a factor of two of three larger than those derived from forbidden
lines of PNe (see Table 1). This result has to be taken into account
in the determination of heavy element abundances from collisionally
excited lines in the IR.

The abundances derived from O recombination lines are typically from 0.2 to 0.3 dex
higher than those derived from O collisionally excited lines and the adoption
of $T_e(4363/5007)$ under the assumption of $t^2$ = 0.00.

For galactic \ion{H}{2} regions the abundances derived from recombination
lines by Esteban et al. (2005) are about 0.2 dex smaller than those derived by 
Deharveng et al. (2000)
and Pilyugin, Ferrini, \& Shkvarun (2003) from O collisionally excited 
lines and the adoption
of $T_e(4363/5007)$ under the assumption of $t^2$ = 0.00.

We present for the first time a calibration of Pagel's method to derive
the O/H ratio in the
8.2 $<$ 12 + log O/H $<$ 8.8 range based on recombination lines. This
calibration is about 0.21 dex higher than that based on
$T_e(4363/5007)$ under the assumption of $t^2$ = 0.00. Alternatively,
the O recombination lines calibration is in fair agreement
with those calibrations based on grids of models that fit the $R_{23}$ 
observed values
like the one by McGaugh (1991); the main reason for the better agreement
is that the 5007[\ion{O}{3}]/H$\beta$
and 3737[\ion{O}{2}]/H$\beta$ ratios depend to a considerably lesser extent on the
$t^2$ value than the 4363/5007 ratio (see Peimbert \& Costero 1969 and
Peimbert et al. 2004).

\acknowledgements

We are grateful to Leticia Carigi, Cesar Esteban, Jorge Garc\'{\i}a-Rojas, 
Mar\'{\i}a Teresa Ruiz, Evan Skillman, and Silvia Torres-Peimbert for many
discussions related to this paper. We are also grateful to Gordon
McAlpine for the magnificent hospitality during this conference.


\begin{thebibliography}

\bibitem{}
Allende-Prieto, C., Barklem, P. S., Lambert, D. L., \&
Kunha, K. 2004,
\aap, 420, 183

\bibitem{}
Asplund, M., Grevesse, N., \& Sauval, A. J. 2005,
in: Cosmic Abundances as Records of Stellar Evolution and
Nucleosynthesis, ed. F. N. Bash \& T. G. Barnes, ASP Conference
Series, in press, astro-ph/0410214

\bibitem{}
Bensby, T., \& Feltzing, S. 2005, in: From Li to U: Elemental
Tracers of Early Cosmic Evolution ed. V. Hill, P. Francois, \& 
F. Primas, IAU Symposium 228, in press, astro-ph/0506145

\bibitem{}
Campbell, A. 1988, 
\apj, 335, 644

\bibitem{}
Carigi, L. , Peimbert, M., Esteban, C., \& Garc\'{\i}a-Rojas, J. 2005,
\apj, 623, 213 

\bibitem{}
Castellanos, M., D\'{\i}az, A. I., \& Terlevich, E. 2002,
\mnras, 329, 315

\bibitem{} 
Deharveng, L., Pe\~na, M., Caplan, J., \& Costero, R. 2000, 
\mnras, 311, 329

\bibitem{}
Dopita, M. A., \& Evans, I. N. 1986,
\apj, 307, 431

\bibitem{}
Edmunds, M. G., \& Pagel, B. E. J. 1984,
\mnras, 211, 507

\bibitem{}
Esteban, C., Garc\'{\i}a-Rojas, J., Peimbert, M., Peimbert, A., Ruiz, M. T., 
Rodr\'{\i}guez, M., \& Carigi, L. 2005,  
\apj, 618, L95

\bibitem{} 
Esteban, C., Peimbert, M., Garc\'{\i}a-Rojas, J., Ruiz, M. T., 
Peimbert, A., \& Rodr\'{\i}guez, M. 2004, 
\mnras, 355, 229

\bibitem{}
Esteban, C., Peimbert, M., Torres-Peimbert, S., \& Rodr\'{\i}guez, M. 2002,
\apj, 581, 241

\bibitem{} 
Garc\'{\i}a-Rojas, J., Esteban, C., Peimbert, M., Rodr\'{\i}guez, M., Ruiz, 
M. T., \& Peimbert, A. 2004,
\apjs, 153, 501

\bibitem{} 
Garc\'{\i}a-Rojas, J., Esteban, C., Peimbert, A.,
Peimbert, M., Rodr\'{\i}guez, M., \& Ruiz, M. T. 2005,
\mnras, in press, (astro-ph/0506409)

\bibitem{} 
Kewley, L. J., \& Dopita, M. A. 2002,
\apjs, 142, 35

\bibitem{} 
Kobulnicky, H. A., Kennicutt, R. C., \& Pizagno, J. L.
1999, 
\apj, 514, 544

\bibitem{} 
Kobulnicky, H. A., \& Kewley, L. J. 2004,
\apj, 617, 240
 
\bibitem{}
Luridiana, V., Peimbert, M., \& Leitherer, C. 1999,
\apj, 527, 110

\bibitem{}
McCall, M. L., Rybski, P. M., \& Shields, G. A. 1985, 
\apjs, 57, 1
 
\bibitem{}
McGaugh, S. S. 1991, 
\apj, 380, 140

\bibitem{}
Pagel, B. E. J., Edmunds, M. G., Blackwell, D. E., Chun, M. S.,
\& Smith G. 1979,
\mnras, 211, 507

\bibitem{}
Peimbert, A. 2003, 
\apj, 584, 735

\bibitem{}
Peimbert, A., Peimbert, M., \& Ruiz, M. T. 2005,
\apj, in press, (astro-ph/0507084)

\bibitem{} 
Peimbert, M. 1967, 
\apj, 150, 825

\bibitem{} 
Peimbert, M., \& Costero, R. 1969,
Bol. Obs. To\-nan\-tzin\-tla y Ta\-cu\-ba\-ya, 5, 3

\bibitem{}
Peimbert, M., \& Peimbert, A. 2003, 
RevMexAASer.Conf., 16, 113 

\bibitem{}
Peimbert, M., Peimbert, A., Ruiz, M. T., \& Esteban, C. 2004,
\apjs, 150, 431, 2004

\adjustfinalcols

\bibitem{}
Peimbert, M., Storey, P. J., \& Torres-Peimbert, S. 1993,
\apj, 414, 626

\bibitem{}
Pilyugin, L. S. 2000,
\aap, 362, 325

\bibitem{}
Pilyugin, L. S. 2003,
\aap, 399, 1003

\bibitem{} 
Pilyugin, L. S., Ferrini, F., \& Shkvarun, R. V. 2003, 
\aap, 401, 557

\bibitem{}
Rela\~no, M., Peimbert, M., \& Beckman, J. 2002,
\apj, 564, 704
 
\bibitem{}
Roy, J.-R., Belley, J., Dutil, Y., \& Martin, P. 1996, 
\apj, 460, 284

\bibitem{}
Ruiz, M. T., Peimbert, A., Peimbert, M., \& Esteban, C. 2003,
\apj, 595, 247

\bibitem{}
Stasi\'nska, G., \& Schaerer, D. 1999,
\aap, 351, 72
 
\bibitem{}
Storey, P. J. 1994,
\aap, 282, 999

\bibitem{}
Torres-Peimbert, S., \& Peimbert, M. 2003, 
in: Planetary Nebulae and Their Role in the Universe, 
ed. S. Kwok, M. Dopita, \& R. Sutherland, IAU 
Symposium 209, 363


\bibitem{} 
Torres-Peimbert, S., Peimbert, M., \& Fierro, J. 1989, 
\apj, 345, 186 

\bibitem{}
Tsamis, Y. G., Barlow, M. J., Liu, X.-W., Danziger, I. J., 
\& Storey, P. J. 2003a,
\mnras, 338, 687

\bibitem{}
Tsamis, Y. G., Barlow, M. J., Liu, X.-W., Danziger, I. J., 
\& Storey, P. J. 2003b,
\mnras, 345, 186

\bibitem{}
Yang, H., Chu, Y.-H., Skillman, E. D., Terlevich, R. 1996,
\aj, 112, 146

\bibitem{}
Zaritsky, D., Kennicutt, R. C., \& Huchra, J. P.1994,
\apj, 420, 87

 
\end{thebibliography}
\end{document}